\documentstyle[12pt]{article}
\newcommand{\newc}{\newcommand} 
\newc{\ra}{\rightarrow} 
\newc{\lra}{\leftrightarrow} 
\newc{\beq}{\begin{equation}} 
\newc{\eeq}{\end{equation}} 
\newc{\barr}{\begin{eqnarray}} 
\newc{\earr}{\end{eqnarray}} 

 1

\begin{document} 
\begin{titlepage}

\begin{center}
{\large \bf Cold Dark Matter in SUSY Theories. The Role of Nuclear Form 
Factors and the Folding with the LSP Velocity}\\

\vspace{15mm} 

T. S. KOSMAS$^a$ \footnote{Present adress: Institute of Theoretical Physics, 
University of T\"ubingen, D-72076, Germany.}
and  J. D. VERGADOS$^{a,b}$ 

\vspace{8mm}

a Theoretical Physics Section, University of Ioannina,\\
 GR 451 10, Ioannina, Greece \\
b Department of Natural Sciences, University of Cyprus,\\
 1678 Nicosia, Cyprus \\
  
\end{center}


\vspace{10mm}

\begin{abstract}
The momentum transfer dependence of the total cross section for elastic
scattering of cold dark matter candidates, i.e. lightest supersymmetric 
particle (LSP), with nuclei is examined. The presented calculations of 
the event rates refer to a number of representative nuclear targets 
throughout the periodic table  and have been obtained in a relatively 
wide phenomenologically allowed SUSY parameter space. 
For the coherent cross sections it is shown that, since the momentum transfer 
can be quite big for large mass of the LSP and heavy nuclei even though 
the energy transfer is small ($\le 100 KeV$), the total cross section can 
in such instances be reduced by a factor of about five. 
For the spin induced cross section of odd-A nuclear targets, as is the case
of $^{207}Pb$ studied in this work, we found that the reduction is less 
pronounced, since the high multipoles tend to enhance the cross section 
as the momentum transfer increases (for LSP mass$\ < $ 200$\ GeV$) and 
partially cancell the momentum retardation.  
The effect of the Earth's revolution around the sun on these event rates 
is also studied by folding with a Maxwellian LSP-velocity distribution 
which is consistent with its density in the halos. We thus found that the 
convoluted event rates do not appreciably change compared to those obtained
with an average velocity. The event rates increase with A and, in the 
SUSY parameter space considered, they can reach values up to 140 
$y^{-1}Kg^{-1}$ for $Pb$. The modulation effect, however, was found to 
be small (less than $\pm 5\%$).
\end{abstract}

\vspace{1.7cm}


\vspace{3.0cm}

\end{titlepage}
\centerline{\bf I.  INTRODUCTION }
\bigskip

There is ample evidence that about $90\%$ of the matter of the universe is
dark \cite{Smith}-\cite{Frenk}. There are, also, numerus arguments 
indicating that our galaxy is immersed in a dark halo which outweights
the luminous component by a factor of about ten.
Furthermore, the large scale structure of the universe, may be 
accommodated supposing two kinds of dark matter \cite{Frenk}.
One kind is composed of particles
which were relativistic at the time of the structure formation. This is called
Hot Dark Matter (HDM). The other kind is composed of particles which were
non-relativistic at the time of structure formation. These constitute the Cold
Dark Matter (CDM) component of the universe. 
In any case the CDM component of the universe is at least 60\%. Obviously,
the galactic halo is composed of dark matter, since HDM particles will
be moving too fast to be trapped in galaxy. There are two candidates
for CDM. The first is MACHO's (Massive Compact Halo Objects, i.e.
white dwarfs, Juppiter-like objects etc.) and the other is more exotic, i.e. 
WIMP's (Weak Interacting Massive Particles). Recent preliminary experiments
\cite{COBE} suggest that about half of the mass of the halo is made
of WIMP's. The most appealing possibility linked with supersymmetry
(SUSY) of a WIMP candidate is the LSP, i.e. the Lightest Supersymmetric 
Particle \cite{Kane}-\cite{Jungm} (see Ref. \cite{Jungm} for a recent review).

In recent years, the phenomenological implications of Supersymmetry are
being taken very seriously \cite{Kane,Barge}. More or less, accurate predictions 
at low energies are now 
feasible in terms of few input parameters in the context of SUSY models
\cite{Kane}-\cite{Leon}. 
Such predictions do not appear to depend on arbitrary choices of the 
relevant parameters or untested assumptions. In any case the SUSY parameter
space is somewhat restricted \cite{Kane}-\cite{Jungm}. 

In such theories derived from Supergravity the LSP is expected to be a
neutral Majorana fermion with mass in the $10-500 GeV/c^2$ region travelling 
with non-relativistic velocities ($<\beta> \approx 10^{-3}$), i.e. with 
energies in the KeV region. 
In practice, however, one expects a velocity distribution which is supposed
to be Maxwellian (see sect. IV). In the absence of
R-parity violating interactions this particle is absolutely stable. But, even
in the presence of R-parity violation, it may live long enough to be a CDM
candidate.

The detection of the LSP, which is going to be denoted by $\chi_1$, is extremely
difficult, since this particle interacts with matter extremely weakly. One
possibility is the detection of secondary high energy neutrinos which are 
produced by pair annihilation in the sun where this particle is trapped.
Such high energy neutrinos can be detected via neutrino telescopes. 

The other possibility, to be examined in the present work, is the detection of
the  energy of the recoiling nucleus $(A,Z)$ in the reaction

\beq
\chi_1  + (A,Z)  \ra \chi_1  + (A,Z)^*
\label{1.1}  
\eeq

\noindent  
This energy can be converted into phonon energy and detected by a temperature
rise in cryostatic detector with sufficiently high Debye temperature 
\cite{Frenk,Cline,Feil}.
The detector should be large enough to allow a sufficient number of counts but
not too large to permit anticoincidence shielding to reduce background.
A compromise of about $1Kg$ is achieved. 

Another possibility is the use of
superconducting granules suspended in a magnetic field. The heat produced
will destroy the superconductor and one can detect the resulting magnetic flux.
Again a target of about $1Kg$ is favored.

There are many targets which can be employed. The most popular ones contain the
nuclei 
$^3_2He$, $^{19}_{9}F$, $^{23}_{11}Na$, $^{29}_{14}Si$,
$^{40}_{20}Ca$, $^{73,}$$^{74}_{32}Ge$,
$^{75}_{33}As$, $^{127}_{53}I$,
$^{134}_{54}Xe$ and $^{207}_{82}Pb$.

In order to be able to calculate the event rate for the process (1)
the following ingredients are necessary.

1) One must be able to construct the effective Lagrangian at the 
elementary particle level in the framework of Supersymmetry 
\cite{Goodm}-\cite{JDV}. We will follow the procedure adopted
in Ref. \cite{JDV}. For the readers convenience we will provide
the important elements in sect. II.

2) One must make the transition from the quark to the nucleon level
\cite{Ashm}-\cite{Chen89}. This is not straightforward for the 
scalar couplings, which dominate the coherent part of the cross section,
and the isoscalar axial current which is important for the incoherent
cross section for odd targets.

3) One must properly treat the nucleus. Admittedly, the uncertainties here 
are smaller than those of even the most restricted SUSY parameter space. 
One, however, would like to put as accurate nuclear physics input as 
possible in order to constrain the SUSY parameters as much as possible
when the data become available.

For the coherent production, if one ignores the momentum transfer
dependence, the procedure is straightforward. The spin matrix element,
however, is another story. For its evaluation practically every
known nuclear model has been employed. 

At first, the Independent Single Particle Shell Model (ISPSM) had been
employed \cite{Goodm,El-Fl,Druck,Freez}.
Subsequent calculations using the Odd Group Method
(OGM) and the Extended Odd Group Method (EOGM), utilizing magnetic
moments and mirror $\beta$-decays, by Engel and Vogel \cite{Eng-Vog},
showed that the ISPSM was inadequate (see also Ref. \cite{El-Fl}). 
Eventually, however, by performing shell model
calculations \cite{Pache}, this model was also found lacking (see Ref.
\cite{Eng-PV}). Iachello, Krauss and Maino \cite{Iache} employed the
Interacting Boson Fermion Model (IBFM) and Nikolaev and Klapdor-
Kleingrothaus \cite{Bedny,Nikol} the finite fermion theory in order to
reliably evaluate the spin matrix elements.

One additional complication arise from the fact that the LSP
appears to be quite massive, perhaps heavier than 100$GeV$.
For such heavy LSP and sufficiently heavy nuclei, the dependence
of the nuclear matrix elements on the momentum transfer cannot
be ignored even if the LSP has energies as low as $100KeV)$ 
\cite{Gries,Druck,Freez}. This affects both the coherent and the spin 
matrix elements.
For the coherent mode the essential new features can be absorbed
in the nuclear form factor. The evaluation of the spin matrix elements is 
quite a bit more complicated. Quite a number of high multipoles can now
contribute, some of them getting contributions from components of
the wave function which do not contribute in the static limit
(i.e. at q=0, see sect. III). Thus, in general, sophisticated Shell Model
calculations are needed to account both for the observed retardation
of the static spin matrix element and its correct q-dependence.
for the experimentally interesting nuclear systems $^{29}_{14}Si$ and
$^{73}_{32}Ge$ very elaborate calculations have been performed  by Ressell 
{\it et al.} \cite{Ress}. In the case of $^{73}_{32}Ge$ 
a further improved calculation by Dimitrov, Engel and Pittel has recently
been performed \cite{Dimit} by suitably mixing variationally determined 
triaxial Slatter determinants. Indeed, for this complex nucleus many 
multipoles contribute and the needed calculations involve techniques which 
are extremely sophisticated.

From the above discussion the necessity for more detailed calculations 
especially for the spin component of the cross sections for heavy nuclei
is motivated.
The aim of the present paper is to calculate LSP-nucleus scattering
cross section using some representative input in the 
restricted SUSY parameter space as outlined above. The coherent matrix
elements are computed throughout the periodic table (sect. III). The needed
form
factors were obtained using the method of Ref. \cite{KV92}, which are in
good agreement with experiment. For the spin matrix elements we have
chosen $^{207}_{82}Pb$ as target. This target, in addition to its 
experimental qualifications, has the advantage of a rather simple nuclear
structure \cite{Ver71} (see sect. III). Thus, only two multipoles can
contribute. 

Finally, the counting rates are folded with a reasonable LSP-velocity 
distribution \cite{Jungm} (see sect. IV) in order to estimate the modulation 
due to the Earth's velocity. Convoluted rates are obtained (sect. IV) in 
the framework of models discussed in the introduction and sect. II.

\bigskip
\bigskip
\centerline{\bf II.  BRIEF DESCRIPTION OF THE OPERATORS}
\bigskip

It has recently been shown that process (\ref{1.1}) can be described by a 
four fermion interaction 
\cite{Goodm}-\cite{Engel} of the type \cite{JDV}

\beq
L_{eff}  = -\frac{G_F}{\sqrt{2}} \, \big[ J_{\lambda} {\bar \chi_1}
\gamma^{\lambda}  \gamma^{5} \chi_1  \, +\,  J {\bar \chi_1}  \chi_1 \big]
\label{1.2}  
\eeq

\noindent  
where

\beq
 J_{\lambda}  = {\bar N} \gamma_{\lambda} [\, f^0_V + f^1_V \tau_3
+( f^0_A + f^1_A \tau_3 ) \gamma_{5} \, ] N  
\label{1.3}  
\eeq

\noindent  
and

\beq
 J  = {\bar N} ( f^0_S + f^1_S \tau_3 ) N
\label{1.4}  
\eeq

\noindent  
We have neglected the uninteresting pseudoscalar and tensor
currents. Note that, due to the Majorana nature of the LSP, 
${\bar \chi_1} \gamma^{\lambda}  \chi_1 =0$ (identically).
The vector and axial vector form factors can arise out of Z-exchange and s-quark
exchange \cite{Goodm}-\cite{JDV}
(s-quarks are the SUSY partners of quarks with spin zero).
They have uncertainties in them (for three choices in the
allowed parameter space of Ref. \cite{Kane} see Ref. \cite{JDV}).
In our choice of the parameters the LSP is mostly a gaugino. Thus, the Z-
contribution is small. It may become dominant in models in which the LSP
happens to be primarily a Higgsino. Such a possibility will be examined 
elsewhere. 
The transition from the quark to the nucleon level is pretty straightforward 
in this case. This is in general the case of vector current contribution. 
We will see later that, due to the Majorana nature of the LSP, the
contribution of the vector current, which can lead to a coherent effect of all
nucleons, is suppressed [10-16]. 
The vector current is effectively multiplied by a factor of $\beta=v/c$, 
$v$ is the velocity of LSP (see Table I).
Thus, the axial current, especially in the
case of light and medium mass nuclei, cannot be ignored.

For the isovector axial current one is pretty confident about how to go 
from the quark to the nucleon level. We know from ordinary weak decays
that the coupling merely gets renormalized from $g_A=1$ to $g_A=1.24$.
For the isoscalar axial current the situation is not completely clear.
The naive quark model (NQM) would give a renormalization parameter of unity
(the same as the isovector vector current). This point of view has, 
however, changed in recent years due to the so-called spin crisis, i.e.
the fact that in the EMC data \cite{Ashm} it appears that only a small fraction
of the proton spin arises from the quarks. Thus, one may have to
renormalize $f^0_A$ by $g^0_A=0.28$, for u and d quarks, and $g^0_A=-0.16$
for the strange quarks \cite{Jaffe}, i.e. a total factor of 0.12. 
These two possibilities, labeled as NQM and EMC, are listed in Table I.
One cannot completely rule out the possibility that the actual value maybe 
anywhere in the above mentioned region \cite{Gensin}.

The scalar form factors arise out of the Higgs exchange or via s-quark 
exchange when there is mixing between s-quarks ${\tilde q}_L$ and 
${\tilde q}_R$ \cite{Goodm}-\cite{El-Fl}
(the partners of the left-handed and right-handed quarks). They have two types 
of uncertainties in them. One, which is the most important,
at the quark
level due to the uncertainties in the Higgs sector. The other in going from  
the quark to the nucleon level \cite{Dree,Engel}.
Such couplings are proportional to the
quark masses, and hence sensitive to the small admixtures of $q {\bar q }$
(q other than u and d) present in the nucleon. Again values of $f^0_S$ and
$f^1_S$ in the allowed SUSY parameter space are considered \cite{JDV}.

The actual values of the parameters $f^0_S$ and $f^1_S$ used here, arising 
mainly from Higgs exchange, were obtained by considering 1-loop corrections
in the Higgs sector. As a result, the lightest Higgs mass is now a bit higher,
i.e. more massive than the value of the Z-boson \cite{El-Ri,Haber}.
The thus obtained values of the parameters $f^0_S$ and $f^1_S$ are smaller
than those of Ref. \cite{JDV} (see Table I). The next source of ambiquities
involves the step of going from the quark to the nucleon level for
the scalar and isoscalar couplings. Here we adopt the procedure
described in Ref. \cite{JDV} as a result of the analysis of Ref.
\cite{Dree,Chen88,Chen89}.

\bigskip
\bigskip
\centerline{\bf III.  TOTAL CROSS SECTION}
\bigskip

The invariant amplitude in the case of non-relativistic LSP takes the form
\cite{JDV}

\barr
|{\it m}|^2 &=& \frac{E_f E_i -m_1^2 +{\bf p}_i\cdot {\bf p}_f } {m_1^2} \,
|J_0|^2 +  |{\bf J}|^2 +  |J|^2 
 \nonumber \\ & \simeq & \beta ^2 |J_0|^2 + |{\bf J}|^2 + |J|^2 
\label{2.1}
 \earr

\noindent  
where $m_1$ is the LSP mass, $|J_0|$ and $|{\bf J }|$ indicate the matrix 
elements of the time and space components of the current $J_\lambda$ 
of Eq. (\ref{1.2}), respectively, and $J$ represents the matrix element of the 
scalar current J of Eq. (\ref{1.3}). Notice that $|J_0|^2$
is multiplied by $\beta^2$ (the suppression due to the Majorana nature of LSP
mentioned above). 
It is straightforward to show that 

\beq
 |J_0|^2 = A^2 |F({\bf q}^2)|^2 \,\left(f^0_V -f^1_V \frac{A-2 Z}{A}
 \right)^2
\label{2.2}
\eeq

\beq
 J^2 = A^2 |F({\bf q}^2)|^2 \,\left(f^0_S -f^1_S \frac{A-2 Z}{A}
 \right)^2
\label{2.3}
\eeq

\beq
 |{\bf J}|^2 = \frac{1}{2J_i+1} |<J_i ||\, [ f^0_A {\bf \Omega}_0({\bf q})
\, + \, f^1_A {\bf \Omega}_1({\bf q}) ] \, ||J_i>|^2 
\label{2.4}
\eeq
with $F({\bf q}^2)$ the nuclear form factor and

\beq 
{\bf \Omega}_0({\bf q})  = \sum_{j=1}^A {\bf \sigma}(j) e^{-i{\bf q} \cdot
{\bf x}_j }, \qquad
{\bf \Omega}_1({\bf q})  = \sum_{j=1}^A {\bf \sigma} (j) {\bf \tau}_3 (j)
 e^{-i{\bf q} \cdot {\bf x}_j }
\label{2.5}  
\eeq

\noindent 
where ${\bf \sigma} (j)$, ${\bf \tau}_3 (j)$, ${\bf x}_j$ are the spin, third
component of isospin ($\tau_3 |p> = |p>$) and coordinate of the j-th nucleon and 
$\bf q$ is the momentum transferred to the nucleus.

The differential cross section in the laboratory frame takes the form 
\cite{JDV}

\barr
\frac{d\sigma}{d \Omega} &=& \frac{\sigma_0}{\pi} (\frac{m_1}{m_p})^2
\frac{1}{(1+\eta)^2} \xi  \{\beta^2 |J_0|^2  [1 - \frac{2\eta+1}{(1+\eta)^2}
\xi^2 ] + |{\bf J}|^2 + |J|^2 \} 
\label{2.6}
 \earr

\noindent 
where $m_p$ is the proton mass, $\eta = m_1/m_p A$, $ $
$\xi = {\bf {\hat p}}_i \cdot {\bf {\hat q}} \ge 0$ (forward scattering) and  

\beq
\sigma_0 = \frac{1}{2\pi} (G_F m_p)^2 \simeq 0.77 \times 10^{-38}cm^2 
\label{2.7} 
\eeq

\noindent 
The momentum transfer  $\bf q$ is given by

\beq
|{\bf q}| = q_0 \xi, \qquad q_0 = \beta \frac{2  m_1 c }{1 +\eta}
\label{2.8} 
\eeq

Some values of $q_0$ (forward momentum transfer) for some characteristic values
of $m_1$ and representative nuclear systems (light, medium and heavy)
are given in Table II. It is clear from Eq. (\ref{2.8}) that the momentum
transfer can be sizable for large $m_1$ and heavy nuclei ($\eta$ small).

The total cross section can be cast in the form

\barr
\sigma &=& \sigma_0 (\frac{m_1}{m_p})^2 \frac{1}{(1+\eta)^2} \,
 \{ A^2 \, [[\beta^2 (f^0_V - f^1_V \frac{A-2 Z}{A})^2 
\nonumber \\ & + & 
(f^0_S - f^1_S \frac{A-2 Z}{A})^2 \, ]I_0(q^2_0) -
\frac{\beta^2}{2} \frac{2\eta +1}{(1+\eta)^2}
(f^0_V - f^1_V \frac{A-2 Z}{A})^2 I_1 (q^2_0) ]
\nonumber \\ & + & 
(f^0_A \Omega_0(0))^2 I_{00}(q^2_0) + 2f^0_A f^1_A \Omega_0(0) \Omega_1(0)
I_{01}(q^2_0) + (f^1_A \Omega_1(0))^2 I_{11}(q^2_0) \, \} 
\label{2.9}
 \earr

\noindent 
The quantities entering eq. (\ref{2.9}) are explained in the subsections
A and B below.

\bigskip
\centerline{\bf A. The coherent matrix element}
\bigskip

The terms in the square bracket of Eq. (\ref{2.9}), describe the coherent
cross section the momentum dependence of which is involved in the integrals 
$I_{\rho}(q^2_0)$, where $\rho$ =0 for the isoscalar and $\rho=1$ for the
isovector component. These integrals are given by

\beq
I_\rho(q^2_0) = 2(\rho +1) \int_0^1 \xi^{1+2\rho} \, |F(q_0^2\xi^2)|^2 \,d\xi,
\qquad \rho = 0,1
\label{2.10} 
\eeq

The integrals $I_{\rho}$ are normalized so as $I_{\rho}(0)=1$ (following the
normalization $F(0)=1$) and they can be calculated by using Ref. \cite{KV92} 
where we have shown that, the nuclear form factor $F(q^2)$ can be
adequately described within the harmonic oscillator model as 

\beq
F({ q^2}) = \,[ \,
\frac{Z}{A} \Phi(qb,Z) + \frac{N}{A} \Phi(qb,N)\, ]\, e^{-q^2b^2/4}
\label{2.14} 
\eeq
$( N = A-Z )$ where $b \approx 1.0 A^{1/6} \, fm$ and $\Phi$ is a polynomial of the form 

\beq
\Phi(qb,\alpha)  = \sum_{\lambda =0}^{N_{max}(\alpha)} \theta_\lambda^{(\alpha)}
(qb)^{2\lambda}, \qquad \alpha = Z, N.
\label{2.15} 
\eeq
$N_{max}(Z)$ and $N_{max}(N)$ depend on the major harmonic oscillator shell
occupied by protons and neutrons \cite{KV92}, respectively. 
The integrals $I_\rho(q^2_0)$ can be written as

\beq
I_\rho(q^2_0)  =  I_\rho(u)  = (1+\rho)u^{-(1+\rho)}
  \int_0^u x^{1+ \rho} \, |F( 2x/b^2)|^2 \,dx,
\label{2.16} 
\eeq
where 

\beq
u = q_0^2b^2/2
\label{2.17} 
\eeq
With the use of Eqs. (\ref{2.14}), (\ref{2.15}) we obtain 

\beq
I_\rho(u)  = \frac{1}{A^2} \{ \, Z^2 I^{(\rho)}_{ZZ}(u) +
2NZI^{(\rho)}_{NZ}(u) +N^2I^{(\rho)}_{NN}(u) \}
\label{2.18} 
\eeq
where

\beq
I^{(\rho)}_{\alpha \beta}(u)  =  (1 + \rho) 
\sum_{\lambda =0}^{N_{max}(\alpha)}  \, \sum_{\nu =0}^{N_{max}(\beta)} 
\frac{\theta_\lambda^{(\alpha)}}{\alpha} \, 
\frac{\theta_\nu^{(\beta)}}{\beta} \,
\frac{2^{\lambda +\nu+\rho} \,(\lambda +\nu+\rho)!}{u^{1+\rho}}
\Big[ 1 - e^{-u} \sum_{\kappa =0}^{\lambda +\nu+\rho} \,\, 
\frac{u^\kappa}{\kappa!} \Big]
 \label{2.19} 
\eeq
($\alpha,\beta = N, Z$).
The coefficients $\theta_\lambda^{(\alpha)}$ are given in Ref. \cite{KV92},
for light and medium nuclei, and in Ref. \cite{KVcdm} for heavy nuclei.

The integrals $I_0$ for three typical nuclei
($^{40}_{20}Ca$, $^{72}_{32}Ge$ and $^{208}_{82}Pb$ ) are presented 
as a function of $m_1$ in Fig. 1. 
The values of the harmonic oscillator parameter b used are:
b=1.849 $fm$ for $Ca$, b=2.039 $fm$ for $Ge$, and b=2.434 $fm$ for $Pb$.
We see that, for light nuclei the modification
of the cross section by the inclusion of the form factor is small.
For heavy nuclei and massive $m_1$ the form factor has a dramatic effect on the
cross section and may decrease it by a factor of about five.
The integral $I_1$ is even more suppressed but it is less important
(see Ref. \cite{KVcdm}).

\bigskip
\centerline{\bf B.  The spin matrix element}
\bigskip

The other terms in Eq. (\ref{2.9}), describe the spin dependence of the 
cross section in terms of the $I_{\rho\rho^{\prime}}$, with
$\rho,\rho^{\prime} =0,1$. The latter integrals result 
by following the standard procedure of the multipole expansion
of the $e^{-i {\bf q} \cdot {\bf r}}$ in Eq. (\ref{2.5}) and
are defined by

\beq
I_{\rho \rho^{\prime}}(q^2_0) = 2 \int_0^1 \xi \, d\xi \sum_{\lambda,\kappa}
\frac{\Omega^{(\lambda,\kappa)}_\rho( q^2_0\xi^2)}{\Omega_\rho (0)} \,
\frac{\Omega^{(\lambda,\kappa)}_{\rho^{\prime}}( q^2_0\xi^2)}
{\Omega_{\rho^{\prime}}(0)} 
, \qquad \rho, \rho^{\prime} = 0,1
\label{2.11} 
\eeq

\noindent
where we have made the identification

\beq
\Omega^{(0,1)}_\rho = \Omega_\rho(q^2_0\xi^2) = (2J_i+1)^{-\frac{1}{2}} 
< J_f|| \sum_{j=1}^A j_0(q_0\xi r_j) \omega_\rho(j) {\bf \sigma}(j) ||J_i>,
\qquad \rho = 0,1
\label{2.12} 
\eeq

\noindent
with $\omega_0(j)=1$ and $\omega_1(j)=\tau_3(j)$. In general,

\beq
\Omega^{(\lambda,\kappa)}_\rho = (2J_i+1)^{-\frac{1}{2}} 
\sqrt{4\pi} < J_f|| \sum_{j=1}^A \Big[ {\bf Y}^{\lambda}({\hat {\bf r}}_j) 
\otimes {\bf \sigma}(j) \Big]^{\kappa}
j_{\lambda}(q_0\xi r_j) \omega_\rho(j) ||J_i>
\qquad \rho = 0,1
\label{2.13} 
\eeq

\noindent
With the above expressions  we have managed to separate the elementary 
parameters $f^0_A$ and $f^1_A$ from the nuclear parameters. 

We warn the reader that, the integrals of Eq. (\ref{2.11}) are normalized 
to unity as $q \ra 0$, i.e. $I_{\rho \rho^{\prime}}(q_0 = 0) = 1$. 
This normalization is different than that found in the previous literature. 
In the above limit ($q_0=0$) the spin matrix element takes the simple expression

$$
|{\bf J}|^2 = \Big|f^0_A \Omega_0(0) + f^1_A \Omega_1(0)\Big|^2 
$$

The spin matrix elements, unfortunately, depend in general rather sensitively 
on the details of the nuclear structure,
which is included in the integrals $I_{\rho\rho^{\prime}}(q_0^2)$ (see Eq.
(\ref{2.11})). As we mentioned in the introduction,
the first attempt for quantitative description
was based on the Odd Group Method \cite{Eng-Vog}. Subsequent shell model calculations
demonstrated that the OGM was not adequate and showed that more elaborate 
calculations were needed \cite{Ress,Dimit}. Furthermore, since the matrix elements at
$q=0$ are often quenched, the momentum dependence of the matrix elements was 
more important than  it was naively expected. As a matter of fact, one has
to include a lot of configurations to accommodate all multipoles, which 
in complex nuclei like $^{29}Si$ and $^{73}Ge$, result in very large Hilbert
spaces. It will be therefore, a very hard task to substantially improve
the calculations of Ref. \cite{Ress}-\cite{Dimit}.

Among the targets which have been considered for LSP detection, 
$^{207}Pb$ stands out as an important candidate. The spin matrix element
of this nucleus has not been evaluated, since one expects the relative 
importance of the spin versus the coherent mode to be more important
on light nuclei. But, as we have mentioned, the spin matrix element in the 
light systems is quenched. On the other hand, the spin matrix element
of $^{207}Pb$, especially the isoscalar one, does not suffer from
unusually large quenching, as is known from the study of the magnetic 
moment. Thus, we view it as a good theoretical laboratory since:
i) It is believed to have simple structure, one $2p 1/2$ neutron
hole outside the doubly magic nucleus $^{208}Pb$.
ii) Because of its low angular momentum, only two multipoles $\lambda =0$
and $\lambda =2$ with a $J$-rank of $\kappa=1$ can contribute even at large
momentum transfers. One can thus view the information obtained from 
this simple nucleus as complementary to that of $^{73}Ge$, which has very 
complex nuclear structure.

To a good approximation \cite{Ver71} the ground state of the 
$^{207}_{82}Pb$ nucleus can be described as a $2p_{1/2}$ neutron hole 
in the  $^{208}_{82}Pb$ closed shell. Then for $\lambda=0$ one finds

\beq
\Omega_0({\bf q}) \, = \, -(1/\sqrt{3}) F_{2p} ({\bf q}^2), \qquad
\Omega_1({\bf q}) \, = \, (1/\sqrt{3}) F_{2p} ({\bf q}^2)
\label{2.20} 
\eeq
and

\beq
I_{00} = I_{01} = I_{11} = 2 \int_0^1 \xi \, [ F_{2p} (q^2) ]^2 \,d\xi
\label{2.21} 
\eeq
Even though the probability of finding a pure $2p_{1/2}$ neutron hole
in the $\frac{1}{2}^-$ ground state of  $^{207}_{82}Pb$ is greater than 95\%,
the ground state magnetic moment is quenched due to the $1^+$ p-h excitation
involving the spin orbit partners. Hence, we expect a similar suppression
of the isovector spin matrix elements. Thus we write

\barr
|{(1/2)}^->_{gs} \, & = & \, C_0 |(2p_{1/2})^{-1}> \Big[ \,
1 + C_1 |[0i_{11/2} (n) (0i_{13/2})^{-1} (n)] 1^+ >
\nonumber \\ & + & 
C_2 |[0h_{9/2} (p) (0h_{11/2})^{-1} (p) ] 1^+ > + ... \Big]
\label{2.22} 
\earr

Due to angular momentum and parity selection rules, we have $\kappa=1$
and $\lambda =0,2$. Retaining terms  which are at most linear in the
coefficients $C_1$, $C_2$ we obtain

i) $\lambda =0$

\beq
\Omega_0({\bf q}) \, = \, C_0^2 \, \{ F_{2p} (q^2) /\sqrt{3}
 -8  \,[  (7/13)^{1/2} C_1 F_{0i} (q^2) \,
+\, (5/11)^{1/2} C_2 F_{0h} (q^2) ]\, \}
\label{2.23} 
\eeq

\beq
\Omega_1({\bf q}) \, = -\, C_0^2 \, \{ F_{2p} (q^2) /\sqrt{3} 
 - 8  \,[  (7/13)^{1/2} C_1 F_{0i} (q^2) \,
-\, (5/11)^{1/2} C_2 F_{0h} (q^2) ]\, \}
\label{2.24} 
\eeq
where

\beq
 F_{nl} (q^2) = e^{-q^2b^2/4} \sum_{\mu =0}^{N_{max}}
\gamma_\mu^{(nl)} (qb)^{2\mu}
\label{2.25} 
\eeq
The coefficients $\gamma_\mu^{(nl)}$ are given in Ref. \cite{KVcdm}
and the coefficients $C_0$,  $C_1$ and  $C_2$ were obtained by diagonalizing the
Kuo-Brown G-matrix \cite{Kuo,Herli} in a model space of 2h-1p configurations. 
They are given by 

$$
C_0 = 0.973350, \qquad C_1 = 0.005295, \qquad  C_2 = -0.006984
$$
We also find

\beq
\Omega_0(0) \, = \, -(1/\sqrt{3}) (0.95659), \qquad (small \, \, \, quenching)
\label{2.26} 
\eeq

\beq
\Omega_1(0) \, = \, (1/\sqrt{3}) (0.83296), \qquad (sizable \, \, \,
quenching) 
\label{2.27} 
\eeq
The amount of retardation of the total matrix element depends on the values  of
$f^0_A$ and $f^1_A$.

ii) $\lambda =2$.

In this case, in addition to the leading $(2p _{1/2})^{-1}$ configuration, 
the first leading correction to the nuclear matrix element is linear in the 
mixing coefficients $C_{j_1j_2}$ appearing in the expression:

\beq
| (\frac{1}{2})^{-1} > = C_0 \Big\{
| (2p_{1/2})^{-1}(n) > + \sum_{j_1j_2} C_{j_1j_2} | 
(2p_{1/2})^{-1}(n); (j^{-1}_1j_2) J_{12}=1;\frac{1}{2}> \Big\} 
\label{2.28} 
\eeq

\noindent
i.e.

\barr
\Omega_{\rho}^{(2,1)} = && \frac{C_0^2}{\sqrt{2J_i+1}}
\{ 2\, \sum_{j_1j_2} C_{j_1j_2} G(j_1,j_2,\rho)
<n_1l_1| j_2(q_0\xi r) |n_2l_2> 
\nonumber \\
&&+(-1)^{\rho} <(2p_{1/2})^{-1}||T^{k}||(2p_{1/2})^{-1}>\}
\label{2.29} 
\earr

\noindent
where $G(j_1,j_2,\rho)$ are isospin dependent geometrical factors
which can be evaluated by standard techniques.
The radial integrals $<n_1l_1| j_2 (q_0\xi r) | n_2l_2 >$ can be cast in the  
form of Eq. (\ref{2.25}), but they depend on two single particle quantum numbers
(see appendix). The relevant coefficients used in the present work
are given in Ref. \cite{KVcdm}.

Notice, however, that 
unlike the $\lambda =0$ case, many amplitudes can contribute if the 
quadrupole modes coupled to the single hole wavefunction are admixed in 
the ground state of the system.

In the simple model of Ref. \cite{Ver71}, in addition to the $C_1$, 
$C_2$ encountered above, one needs the amplitudes of the two additional
configurations, $0j_{13/2} 0j_{11/2}(n)$ and $0i_{11/2} 0g_{9/2} (p)$,
which are $C_3=0.000239$ and $C_4=-0.000642$. Obviously, this is a 
simplification, since one should consider the spin Giant Quadrupole
Resonance (GQR), which may have a small admixture in the ground state
of the nucleus but a very large transition matrix element. Such a 
detailed calculation including all 2$\hbar \omega$ excitations is
in progress and it will be reported elsewhere.

Using Eqs. (\ref{2.11}), (\ref{2.23}), (\ref{2.24}) and (\ref{2.29}),
we can evaluate the integrals $I_{00}$, $I_{01}$ and $I_{00}$
for $^{207}Pb$. The results for $I_{11}$ are presented in Fig. 2
(the other two are practically indistinguishable, see Ref. \cite{KVcdm}). 
In Fig. 2(a) $I_{11}$ is plotted as a function of the LSP mass while in Fig. 
2(b) it is plotted (together with $I_0$) as a function of the parameter u 
given in Eq. (\ref{2.17}). 
We see that for a heavy nucleus and high LSP mass the momentum transfer 
dependence of the spin monopole $(\lambda=0)$ matrix elements is quite large. 
It is, however, to a large extent neutralized by the spin quadrupole 
$(\lambda=2)$. So the overall effect is not dramatic for LSP mass less than 
$100 GeV$. As a matter of fact, from Fig. 2(b) we see that, the retardation of
the spin matrix element is quite a bit less than that of the coherent mode
for almost all values of u.

\medskip
\bigskip
\centerline{\bf IV.  CONVOLUTION OF THE CROSS SECTION WITH }  
\centerline{\bf THE VELOCITY DISTRIBUTION} 
\bigskip

The  cross sections which would be given from an LSP-detector participating 
in the revolution of the earth around the sun would appear retarded. In this
section we are going to study this effect by using the method of folding. To 
this aim let us assume that the LSP is moving with velocity $v_z$ with 
respect to the detecting apparatus. Then the detection rate 
for a target with mass $m$ is given by

\beq
\frac{dN}{dt} =\frac{\rho (0)}{m_1} \frac{m}{A m_p} | v_z | \sigma (v)
\label{4.1}  
\eeq

\noindent
where $\rho (0) = 0.3 GeV/cm^3$ is the LSP density in our vicinity. 
This density has to be consistent with the LSP velocity distribution.
Such a consistent choise can be a Maxwell distribution

\beq
f(v^{\prime}) = (\sqrt{\pi}v_0)^{-3} e^{-(v^{\prime}/v_0)^2 }
\label{4.2}  
\eeq

\noindent
provided that \cite{Iache}

\beq
v_0 = \sqrt{(2/3)<v^2>} =220 Km /s
\label{4.3}  
\eeq

\noindent
For our purposes it is convenient to express the above distribution in the
laboratory frame, i.e.

\beq
f({\bf v}, {\bf v}_E) = (\sqrt{\pi}v_0)^{-3} 
e^{- ({\bf v}+{\bf v}_E)^2/v_0^2}
\label{4.4}  
\eeq

\noindent
where ${\bf v}_E$ is the velocity of the earth with respect to the center
of the distribution. Choosing a coordinate system in which ${\bf \hat  x}_2$ 
is the axis of the galaxy, ${\bf \hat  x}_3$ is
along the sun's direction of motion (${\bf v}_0$)
and ${\bf \hat  x}_1 = {\bf \hat  x}_2 \times {\bf \hat  x}_3$, 
we find that the position of the axis of the ecliptic is determined
by the angle $\gamma \approx 29.80$ (galactic latitude) and the
azymouthal angle $\omega = 186.3^0$ measured on the galactic plane
from the ${\bf \hat  x}_3$ axis \cite{Alissan}.

Thus, the axis of the ecliptic lies very close to the $x_2x_3$ plane
and the velocity of the earth is

\beq
{\bf v}_E \, = \, {\bf v}_0 \, + \, {\bf v}_1 \, 
= \, {\bf v}_0 + v_1(\, sin{\alpha} \, {\bf \hat x}_1
-cos {\alpha} \, cos{\gamma} \, {\bf \hat x}_2
+cos {\alpha} \, sin{\gamma} \, {\bf \hat x}_3\,)
\label{4.5}  
\eeq

\noindent
and

\beq
{\bf v}_0 \cdot {\bf v}_1 = v_0 v_1 
\frac{cos\, \alpha}{\sqrt{1 + cot^2 \gamma \, cos^2\omega}}
\approx v_0 v_1 \, sin\, \gamma \, cos\,\alpha
\label{4.6}  
\eeq

\noindent
where 
${ v}_0$ is the velocity of the sun around the center of the galaxy,
${ v}_1$ is the speed of the earth's revolution around the sun,
$\alpha$ is the phase of the earth orbital motion, $\alpha =2\pi 
(t-t_1)/T_E$, where $t_1$ is around second of June and
$T_E =1 year$.

The mean value of the event rate of Eq. (\ref{4.1}), is defined by

\beq
\Big<\frac{dN}{dt}\Big> =\frac{\rho (0)}{m_1} 
\frac{m}{A m_p} 
\int f({\bf v}, {\bf v}_E) \mid v_z \mid \sigma (|{\bf v}|)
d^3 {\bf v} 
\label{4.7}  
\eeq

\noindent
Then we can write the counting rate as

\beq
\Big<\frac{dN}{dt}\Big> =\frac{\rho (0)}{m_1} \frac{m}{Am_p} \sqrt{<v^2>}
<\Sigma>
\label{4.8}  
\eeq

\noindent
where

\beq
< \Sigma > =\int \frac{ | v_z | } {\sqrt{<v^2>}} 
f({\bf v}, {\bf v}_E) \sigma (|{\bf v}|) d^3 {\bf v}
\label{4.9}  
\eeq

\noindent
Thus, taking the polar axis in the direction ${\bf v}_E$, we get

\beq
< \Sigma > = \frac{4}{\sqrt{6\pi} v_0^{4}}
\int_0^{\infty} v^3 d v \int_{-1}^{1} |\xi| d \xi 
e^{-(v^2+v_E^2+2v v_E \xi)/v_0^2 } \sigma (v) 
\label{4.10}  
\eeq

\noindent
or

\beq
< \Sigma > = \frac{2}{\sqrt{6\pi} v_E^{2}}
\int_0^{\infty} v d v \, F_0(\frac{2vv_E}{v^2_0}) \,
e^{-(v^2+v_E^2)/v_0^2 } \sigma (v) 
\label{4.11}  
\eeq

\noindent
with 

\beq
F_0(\chi) =\chi sinh \chi - cosh \chi + 1
\label{4.12}  
\eeq

One can also write Eq. (\ref{4.11}) as followes

\beq
< \Sigma > = \Big(\frac{2}{3} \Big)^{\frac{1}{2}}
\int_0^{\infty} \frac{v}{v_0} \, f_1(v) \sigma (v) dv
\label{4.13}  
\eeq

\noindent
with

\beq
f_1(v) \, = \, \frac{1}{\sqrt{\pi}} \, \frac{v_0}{v^2_E}\, 
F_0(\frac{2vv_E}{v^2_0}) \, e^{-(v^2+v_E^2)/v_0^2 }
\label{4.14}  
\eeq

\noindent
In the case in which the first term in Eq. (\ref{4.12}) becomes dominant,
we get 

\beq
f_1(v) \, = \, \frac{1}{\sqrt{\pi}} \, \frac{v}{v_0v_E}\, 
\Big\{ \, exp\Big[\, - \frac{(v-v_E)^2}{v_0^2 }\Big]\, - \,
exp\Big[\, - \frac{(v+v_E)^2}{v_0^2 }\Big]\, \Big\}
\label{4.15}  
\eeq

\noindent
in agreement with Eq. (8.15) of Ref. \cite{Jungm}. In Eq. (\ref{4.11})
the nuclear parameters are implicit in the cross section $\sigma(v)$ 
given from Eq. (\ref{2.9}). The nuclear physics dependence of $<\Sigma>$
could be disentangled by taking note of the extra velocity dependence 
of the coherent vector contribution in $\sigma(v)$
and introducing the parameters

\beq
\delta = \frac{2 v_E }{v_0}\, = \, 0.27,
\qquad  \psi = \frac{ v }{v_0}, \qquad  u = u_0 \psi ^2
\label{4.16}  
\eeq

\noindent
where the quantity $u_0$ is the one entering the nuclear form factors of Eq. 
(\ref{2.16}) for $v=v_0$, which in this case is given by

\beq
u_0 = \frac{1}{2} \left( \frac{2\beta_0 m_1 c^2}{(1+\eta)} 
\frac{b}{\hbar c}\right)^2,
\qquad \beta_0 = \frac{v_0}{c}
\label{4.17}  
\eeq
Afterwards, we can write Eq. (\ref{4.11}) as

\barr
<\Sigma>&=&\Big(\frac{m_1}{m_p}\Big)^2 \frac{\sigma_0}{(1+\eta)^2} \\
\nonumber
&&\Big\{A^2 \Big[ <\beta^2> \Big(f^0_V-f^1_V \frac{A-2 Z}{A})^2 
\Big(J_0-\frac{2\eta+1}{2(1+\eta)^2}J_1\Big) +
(f^0_S-f^1_S\frac{A-2 Z}{A})^2{\tilde J}_0\Big]  \\
\nonumber
&& + \Big( f^0_A \Omega_0(0)\Big)^2 J_{00}
+ 2 f^0_A f^1_A \Omega_0(0)\Omega_1(0) J_{01}
+ \Big( f^1_A \Omega_1(0)\Big)^2 J_{11} \Big\}
\label{4.18}  
\earr

If we assume that $J_{00}=J_{01}=J_{11}$, as seems to be the case for 
$^{207}Pb$, the spin dependent part of Eq. (\ref{4.14}) is reduced 
to the familiar expression 
$\Big[f^0_A \Omega_0(0) + f^1_A \Omega_1(0)\Big]^2 J_{11}$, where the quantity
in the bracket represents the spin matrix element at $q=0$.

The parameters ${\tilde J}_0$, $J_\rho$, $J_{\rho\sigma}$ describe the
scalar, vector and spin part of the counting rate, respectively,
and they are given by

\beq
{\tilde J}_0 (\lambda, u_0)  \,= \, \frac{2}{\sqrt{6\pi}}
\frac{e^{-\lambda^2}}{\lambda^2}
\int_0^{\infty}\psi e^{-\psi^2 } F_0 ( 2 \lambda \psi) I_0 (u_0\psi^2)d \psi
\label{4.19}  
\eeq

\beq
J_{\rho}(\lambda, u_0)  \,= \, \frac{2}{\sqrt{6\pi}}
\frac{e^{-\lambda^2}}{\lambda^2}
\int_0^{\infty}\psi^3 e^{-\psi^2 } F_0 ( 2 \lambda \psi) 
I_{\rho}(u_0\psi^2)d \psi
\label{4.20}  
\eeq

\beq
J_{\rho\sigma}(\lambda, u_0)  \,= \, \frac{2}{\sqrt{6\pi}}
\frac{e^{-\lambda^2}}{\lambda^2}
\int_0^{\infty}\psi e^{-\psi^2 } F_0 ( 2 \lambda \psi) 
I_{\rho\sigma}(u_0\psi^2)d \psi
\label{4.21}  
\eeq

\beq
\lambda \, = \, \frac{v_E}{v_0} = \Big[ \, 1 + \delta cos\alpha
sin\gamma + (\delta/2)^2 \Big]^{1/2}
\label{4.22}  
\eeq

The parameters $I_\rho$,  $I_{\rho\sigma}$ have been discussed in the 
previous section. The above integrals are functions of $\lambda$
and $u_0$. The latter depends on $v_0$, the nuclear parameters
and the LSP mass. These integrals can only be done numerically.
Since, however, $\lambda$ is close to unity, we can expand in powers of
$\delta$ and make explicit the dependence of these integrals on the earth's 
motion. Thus,

\beq
{\tilde J}_0(\lambda, u_0)  \,= \, \frac{2}{\sqrt{6\pi}}
B_1\,\Big[ \, {\tilde K}^{(0)}_0(u_0) 
+ \delta\, sin\gamma\, cos\alpha
 \, {\tilde K}^{(1)}_0(u_0)  \Big]
\label{4.23}  
\eeq

\beq
J_{\rho}(\lambda, u_0)  \,= \, \frac{2}{\sqrt{6\pi}}
B_2\,\Big[ \, K^{(0)}_{\rho}(u_0) 
+ \delta\, sin\gamma\, cos\alpha \, K^{(1)}_{\rho}(u_0)  \Big]
\label{4.24}  
\eeq

\beq
J_{\rho\sigma}(\lambda, u_0)  \,= \, \frac{2}{\sqrt{6\pi}}
B_1 \,\Big[ \,  K^{(0)}_{\rho\sigma}(u_0) 
+ \delta\, sin\gamma\, cos\alpha
 \,  K^{(1)}_{\rho\sigma}(u_0) \Big]
\label{4.25}  
\eeq

The integrals ${\tilde K}^{0}_{0}$, $K^{0}_{\rho}$ and  
$K^{0}_{\rho\sigma}$ are normalized so that they become unity at 
$u_0 =0$ (negligible momentum transfer). We find

\beq
B_1\, = \, \frac{1}{e}
\int_0^{\infty}\psi e^{-\psi^2 } F_0(2\psi) d \psi =
 \, \frac{1}{e} + 2\nu \, \approx \, 1.860
\label{4.26}  
\eeq

\beq
B_2\, = \, \frac{2}{3e}
\int_0^{\infty}\psi^3 e^{-\psi^2 } F_0(2\psi) d \psi
= \, \frac{2}{3} (\, \frac{3}{e} + 7 \nu ) \, \approx \, 4.220
\label{4.27}  
\eeq

\noindent
with

\beq
\nu \, = \, \int_0^{1} e^{-t^2 } d t \,  \approx  \, 0.747
\label{4.28}  
\eeq

\noindent
Furthermore,

\beq
{\tilde K}^{l}_0  \,= \, \frac{1}{eB_1}
\int_0^{\infty}\psi e^{-\psi^2 } F_l(2\psi) I_0 (u_0\psi^2)d \psi,
\qquad l=0,1
\label{4.30}  
\eeq

\noindent

\beq
K^{l}_{\rho} = \frac{2}{3eB_2}\,
\int_0^{\infty}\psi^3 e^{-\psi^2 } F_l(2\psi) I_{\rho} (u_0\psi^2)d \psi,
\qquad l=0,1
\label{4.31}  
\eeq

\noindent

\beq
K^{l}_{\rho\sigma} = \frac{1}{eB_1}\,
\int_0^{\infty}\psi e^{-\psi^2 } F_l(2\psi) I_{\rho\sigma} (u_0\psi^2)d \psi,
\qquad l=0,1
\label{4.32}  
\eeq

\noindent
with $F_0(\chi)$ given in Eq. (\ref{4.12}) and

\beq
F_1(\chi) \, = \, 2\, \Big[ \,(\frac{\chi^2}{4} + 1) cosh\, \chi - 
\chi \,sinh \,\chi -1 \, \Big]
\label{4.33}
\eeq

\noindent
The counting rate can thus be cast in the form

\beq
\Big<\frac{dN}{dt}\Big>=\Big<\frac{dN}{dt}\Big>_{0}(1+hcos\alpha)
\label{4.34}
\eeq

\noindent
where $\big<\frac{dN}{dt}\big>_0$ is the rate obtained from the $l=0$ multipole 
and $h$ the amplitude of the oscillation, i.e. the ratio of the component
of the multipole $l=1$ to that of the multipole $l=0$. Below (see also Tables 
III and IV) we compute separately the amplitude of oscillation for the scalar, 
vector and spin parts of the event rate i.e. the quantity
$h=\delta sin\gamma \, K^{1}(u_0)/K^{0}(u_0)$. 
Note the presence of the geometric factor $sin \gamma = 1/2$, which reduces 
the modulation effect.

In order to get some idea of the dependence of the counting rate on the 
earth's motion, we will evaluate the above expressions at $u_0=0$.
We get

\beq
{\tilde K}^{0}_0  \,= \, 
K^{0}_{\rho\sigma} \approx K^{0}_{\rho}   \,= \, 1
\label{4.35}
\eeq

\beq
{\tilde K}^{1}_0  \,= \, 
K^{1}_{\rho\sigma} \,= \, \frac{\nu}{1/e +2\nu} \approx 0.402
\label{4.36}
\eeq

\beq
K^{1}_{\rho}   
 \,= \, \frac{3/(2e) + (11/2)\nu/2}{3/e +7\nu} \approx 0.736
\label{4.37}
\eeq

\noindent
Thus, for $sin\gamma \approx 0.5$

\beq
{\tilde J}_0 \approx J_{\rho\sigma} = \frac{2}{\sqrt{6\pi}}\, 1.860 
(1 + 0.054 cos\alpha) = 0.857(1+ 0.054 cos\alpha)
\label{4.38}
\eeq

\beq
J_{\rho} = \frac{2}{\sqrt{6\pi}}\, 4.220 (1 + 0.099 cos\alpha)
= 1.944 (1+ 0.099 cos\alpha)
\label{4.39}
\eeq

We see that, the modulation of the detection rate due to the earth's motion
is quite small ($h \approx 0.05$). The corresponding amplitude of 
oscillation in the coherent vector contribution, Eq. (\ref{4.38}), is a bit 
bigger ($h \approx 0.10$). However, this contribution is suppressed due to 
the Majorana nature of LSP (through the factor $\beta^2$). The modulation due 
to the Earth's rotation is expected to be even smaller.

The exact $K^{l}$ integrals, for the l=0 and l=1, are shown in Figs.
3(a)-3(c). The most important of these integrals, those of 
Eq. (\ref{4.30}) associated with the
scalar interaction, are shown in Fig. 3(a). In Fig. 3(b) we present the 
integrals of Eq. (\ref{4.31}) for $\rho=0 $ associated with the vector 
interaction (the integral for  
$\rho=1 $ is analogous but it is less important). Finally in Fig. 3(c) the    
integrals of Eq. (\ref{4.32}) for $\rho=1 $ and $ \sigma=1 $ are shown.
The others are practically indistinquishable from these and are not shown
(see Ref. \cite{KVcdm}). 

Before closing this section we should mention that, the folding procedure can
also be applied in the differential rate in order to obtain the corresponding 
convoluted expression for $d\sigma/d\Omega$, i.e. before doing the angular 
integration in Eq. (\ref{2.6}) and obtain the total cross section Eq. 
(\ref{2.9}). The resulting expressions are, however, a bit more complicated 
and they will not be given here.

\bigskip
\bigskip
\centerline{\bf V.  RESULTS AND DISCUSSION }
\bigskip

The main goal of the present work was the calculation of the cross sections 
for the scattering of LSP, a Cold Dark Matter Candidate, with nuclei.
The coherent scattering was evaluated for three typical nuclei, 
$Ca$, $Ge$ and $Pb$, and the spin matrix element for $^{207}Pb$. 
The momentum dependence of the matrix elements was taken into account.
Special attention has been paid to evaluating
the modulation of the event rates due to the earth's motion.
Our results are summarized as follows.

\bigskip
\centerline{\bf A.  Coherent scattering}
\bigskip

As we have seen in sect. III,
the coherent scattering depends on the isoscalar scalar, $f^0_S$, and vector, 
$f^0_V$, parameters. The latter is effectively multiplied by the average 
velocity $\Big<\beta^2\Big>^{1/2}$ of the LSP due to its Majorana nature.
These parameters depend on the SUSY model considered. They were evaluated 
in the allowed SUSY parameter space of Kane et al. \cite {Kane}. 
The construction of the dominant scalar parameters suffers from additional 
uncertainties, which involve the step of going from the quark to the nucleon 
level. In other words, the results are very sensitive to the presence of 
quarks other than u and d in the nucleon. Three such choices indicated by A, 
B, C are presented in Table I. 
 
From a nuclear structure input point of view, the coherent scattering does 
not depend on the details of the nuclear wave function. It does, however,  
depend on the nuclear density, i.e. the assumed form factor, for fairly massive
LSP and correspondingly heavy nuclei. This is because the momentum transfer in
such cases can be quite high (by nuclear standards) even though the energy 
transfer is small. The form factors used were as realistic as possible 
\cite{KV92}. The inclusion of the form factor results in sizable retardation 
of the cross section which, for case \#1, can be a factor of 3 for Ge, and 
a factor of 14 for Pb. This conclusion is in agreement with previous estimates 
although they do not include detailed calculations for heavy nuclear targets
\cite{Ress}.

\bigskip
\centerline{\bf B. Spin Contribution}
\bigskip

As we have  
mentioned in the introduction, the relative importance of the spin matrix 
element is expected to decrease as one goes to the heavier systems vis a vis 
the coherent scattering. We have seen, however, that the increase due to the
mass number in the coherent scattering is offset by the decline due to form 
factor. Thus the spin contribution may not be completely negligible.

For the spin matrix contribution of $^{207}Pb$, we find that its 
ground state (a $\frac{1}{2}^-$ state), in a 1h and 2h-1p model 
space, is more than 95\% a $2p_{1/2}$ neutron configuration. We have evaluated 
the spin  matrix elements up to terms linear in the small amplitudes. Out of
the 250 components, only two of them contribute to the $\lambda=0$ multipole 
while two more contribute to the $\lambda=2$ multipole. Thus, we find that, the 
isoscalar static matrix element suffers from very little retardation, while 
the isovector matrix element is reduced by 17\%. Since, in the SUSY parameter 
space we have used, the isovector coupling $f^1_A$ is larger, this results 
in a similar retardation of the total matrix element. The isoscalar matrix 
element has other uncertainties. In passing from the quark to the nucleon 
level the amplitude must be multiplied by a factor $g_A$ which ranges from 1,
in the Naive Quark Model (NQM), to 0.12, if extracted from the EMC data. 
In our calculations we considered both of these extremes. Also, since the 
dominant configuration is of the neutron variety, the isoscalar tends to 
subtract from the isovector, but this is not so dramatic in our case since 
the isoscalar is smaller in absolute value, especially in the EMC case.

Our spin matrix elements at $q=0$ are 

$$
\Big[\Omega_1(0)\Big]^2 \, = \, 0.231 \qquad and \qquad
\Big[\Omega_0(0)\Big]^2 \, = \, 0.301
$$

\noindent
These are a bit larger than the values

$$
\Big[\Omega_1(0)\Big]^2 \,  \approx \,
\Big[\Omega_0(0)\Big]^2 \, = \, 0.2 \qquad (\frac{1}{2}^+ \,\,\, ^{29}Si)
$$

\noindent
extracted from Fig. 3 of Ref. \cite{Ress}. They are, however, smaller than 
the values

$$
\Big[\Omega_1(0)\Big]^2 \, =  \, 1.00 \qquad and
\qquad \Big[\Omega_0(0)\Big]^2 \, = \, 1.03 \qquad (\frac{9}{2}^+ \,\,\, 
^{73}Ge)
$$

\noindent
extracted from Fig. 3 of Ref. \cite{Ress}. 

As it has been found in Eq. (\ref{2.9}),
the momentum dependence of the spin part
can be described in terms of three integrals $I_{00}$, 
$I_{01}$, $I_{11}$, where the subscripts indicate the isospin channels. In
our case, these integrals receive contributions from two multipoles, 
$\lambda=0$ (spin monopole) and $\lambda=2$ (spin-quadrupole).
They were judiciously normalized to unity at $q=0 $. 
When so normalized these integrals are approximately equal (see Fig. 3 and
also Ref. \cite{KVcdm}).
We notice that, the monopole contribution falls 
with momentum transfer, as expected, quite fast in fact for large LSP mass.
On the contrary, the quadrupole contribution starts out at zero and keeps 
increasing up to about 90 GeV. As a result its contribution in this mass regime 
is crucial, since it tends to partly compensate for the suppression of the 
monopole term. We expect these trends to persist even in the most elaborate
calculations which include the Giant Quadrupole Resonance (GQR) and are 
currently under way.

As a consequence the momentum transfer suppression is small 
(a factor of about 4 for $m_1=100 GeV$). 
Thus, the above two matrix elements take the values
0.120 and 0.153 for the purely isovector and isoscalar channels. From Table
IV of Ref. \cite{Ress} one extracts at $y$=0.174 the values 0.335 and 0.402.
Thus the spin matrix element 
$\Omega^2$ in $^{207}Pb$ is less than three times smaller 
compared to that of Ref. \cite{Ress} for $^{73}Ge$. On the other hand the
corresponding matrix elements for the $A=29$ and $A=207$ are similar.
When it comes to
the cross section the spin contribution in the case of the $A=207$ 
and $A=73$ may 
be similar due to the presence of $\eta$ in Eq. (\ref{2.9}), which favors 
the $A=207$ system by a factor of about 2.6 for $m_1=100 GeV$.

\bigskip
\bigskip
\centerline{\bf C.  Convoluted rates}
\bigskip

For the coherent cross section in Table III we present results for the
parameters $\Big<\frac{dN}{dt}\Big>_0$ and h for three typical and 
experimentally interesting nuclei, $Ca$, $Ge$ and $Pb$. 
From this table we see that, the event rate is highest when the 
intermediate particles and the LSP are lightest ($m_1=27GeV$, case \#2). 
We notice that, even within the allowed parameter space, the event rate 
may vary by two orders of magnitude. We also notice that,
with the possible exception of the not so realistic model A,
the Higgs contribution becomes dominant.
This is even more so in models where quarks other than u and d are present 
in the nucleus with appreciable probabilities, due to their large masses. 
In the most favorable cases, one may have more than 140 events per year per
kilogram of $Pb$ target. We notice that the amplitude due to the Earth's annual
motion depends on $m_1$ and $A$. In $Pb$ it is less than $\pm 5\%$.

Finally, we should mention that, for cases \#1 
and \#3 (massive LSP), the rate due to the nuclear form factor can be reduced 
by a factor of approximately 6. Because the form factor dependence is 
more pronounced in heavy nuclei, if the LSP turns out to be massive,
the benefit of going to heavier targets is somewhat diminished.

The detection rates obtained, after folding the spin induced cross section,
for $^{207}Pb$ are shown in Table IV. 
The folding makes the spin dependence even less sensitive on
the momentum transfer. The event rate, however, remains  much smaller than the 
coherent scattering for models B and C and even for model A in the case \#2. 
These results may suggest that, in the SUSY 
parameter space given in Ref. \cite{Kane}, the spin induced LSP-nucleus
scattering may not be detectable.     
The spin contribution, however, maybe somewhat enhanced in models in which
the LSP is primarily a Higgsino \cite{KA-JA}.

In order to estimate the total effect of the nuclear density on the 
event rates, in Tables III and IV the results by ignoring the nuclear
form factor (WNFF) are shown.

\bigskip
\bigskip
\centerline{\bf VI.  SUMMARY  AND  CONCLUSIONS}
\bigskip

In the present work we have performed calculations of the event rate for
LSP-nucleus scattering for three typical experimentally interesting nuclei,
i.e.  Ca, Ge  and Pb. The three basic ingredients of our calculation were 
the input SUSY parameters, a quark model for the nucleon and the structure
of the nuclei involved.

The input SUSY parameters were calculated in a phenomenologically allowed 
parameter space (cases \#1, \#2, \#3 of Table I) as explained in the text.
In going from the quark to the nucleon level, the quark structure of the 
nucleon was essential, in particular its content in quarks other than u and d.
For the scalar
interaction we considered three models (labeled A, B, C in Table I) as
described in the text. For the isovector axial coupling one encounters
the so-called nucleon spin crisis. Again we considered two possibilities 
depending on the assumed  portion of the nucleon spin which is due to the
quarks (indicated by EMC and NQM in Table I) as described in the text.

As regards nuclear structure, we employed as detailed as feasible nuclear wave
functions. For the coherent part (scalar and vector) we used realistic nuclear
form factors. For the $^{207}Pb$ system we also computed the spin matrix 
element.
The ground state wave function was obtained by diagonalizing the nuclear 
Hamiltonian in a 2h-1p space which is standard for this doubly magic nucleus.   
The momentum dependence of the matrix elements was taken into account and all 
relevant multipoles were retained (in this system one encounters only two, 
$\lambda=0$ and $\lambda=2$, due to selection rules).

From our discussion in the previous section we conclude that, even though
the spin matrix elements $\Omega^2$ are a factor of 3 smaller than those
for $^{73}Ge$ obtained in Ref. \cite{Ress}, their contribution to the 
cross section is almost
the same (at least for LSP masses around $100 GeV$).

Finally, the obtained results were convoluted with a suitable Maxwell-Boltzmann
velocity distribution of the LSP's. This convolution was necessary to partially
neutralize the form factor retardation, but mainly to compute the modulation of
the event rate due to the Earth's motion. We find that in almost all cases the
event rate due to the Earth's revolution around the sun is very small, less than
$\pm 5\%$ around its average value. The event rates thus obtained (see Tables 
III and IV) are, unfortunately, sensitive to the input parameters. 

The inclusion of the nuclear form factor significantly retards the event rates 
for heavy nuclei ($A>100$) and fairly massive LSP ($m_1>100 GeV$). However, 
this retardation does not outweigh completely the advantages of using a heavy 
target. For the spin matrix elements the form factor retardation of the usual 
$\lambda=0$ multipole is partially neutralized by the enhancement of the 
$\lambda=2$ multipole.

From the data of Tables III and IV we see that it is possible to have detectable
rates ( $>20$ per kilogram per year) only for case \#2 and the realistic
nucleon models B and C, associated with the scalar Higgs exchange term. 
In all other cases, including the spin contribution, the calculated event 
rates are too small.

\bigskip
\bigskip
\bigskip
\centerline{\bf  ACKNOWLEDGEMENTS }
\medskip

This work was partially supported by the grants PENED91 and PENED95 
of the Greek Secretariat for Research. The hospitality of the University of 
Cyprus to one of us (J.D.V.) is also happily acknowledged.

\bigskip
\centerline{\bf APPENDIX }
\bigskip

For the spin contribution of $^{207}Pb$, we have to calculate the matrix 
elements of the operator 

\beq
T^{\kappa} = \sqrt{4\pi} j_\lambda(qr) \Big[ Y^\lambda({\hat {\bf q}})
\otimes \sigma \Big]^{\kappa}
\label{A.1}  
\eeq

\noindent
where $\kappa =1 $ and $\lambda =0, 2$ for the states (see Eqs. (\ref{2.22}),
(\ref{2.28})) 

\beq
|(2p_{1/2})^{-1}> , \qquad C_{j_1j_2} | \Big[ 2p_{1/2})^{-1} j_1^{-1}\Big]
J; j_2; 1/2 >
\label{A.2}  
\eeq

\noindent
or  the matrix elements

\barr
ME &=& C_0^2 \Big[ \, \pm \, <(2p_{1/2})^{-1}||T^{\kappa}||(2p_{1/2})^{-1}> + \\
\nonumber
&\pm& 2 \sum_{j_1j_2} \, C_{j_1j_2} <(2p_{1/2})^{-1}||T^{\kappa}||
(2p_{1/2})^{-1} [j_1^{-1} j_2 ]\kappa; 1/2 >
\label{A.3}  
\earr

\noindent
The $+$ sign is for isoscalar and the $-$ for isovector matrix elements.
The reduced matrix elements $<j_1||T^{\kappa}||j_2>$ are given in
ref. \cite{KVCF}. The relevant radial matrix elements 
$<n_1l_1| j_l(qr) |n_2l_2>$ for harmonic oscillator 
basis can be witten in the compact way 

\beq
<n_1l_1| j_l(qr) |n_2l_2> \, = \,  e^{-\chi} 
\sum_{\kappa=0}^{\kappa_{max}} \, \varepsilon_{\kappa} \, \chi^{\kappa + l/2}
, \qquad  \chi = (qb)^2/4
\label{A.4}  
\eeq

\noindent
where

$$
\kappa_{max} = n_1 + n_2 +m, \qquad m=(l_1 + l_2 -l)/2, \qquad
| l_1 \, - \,l_2 | \, \le \, l \, \le \, l_1 + l_2
$$

\noindent
The coefficients $\varepsilon_{\kappa}(n_1l_1,n_2l_2,l)$ are given by

\beq
\varepsilon_{\kappa} = \Big[ \frac{\pi \, n_1 ! n_2 !}{4 \,
\Gamma(n_1+l_1+\frac{3}{2}) \Gamma(n_2+l_2+\frac{3}{2})} 
\Big]^{\frac{1}{2}} \, \,
\sum_{\kappa_1 =\phi}^{n_1} \,\, \sum_{\kappa_2 =\sigma}^{n_2} \, n! \,
\Lambda_{\kappa_1}(n_1 l_1) \Lambda_{\kappa_2}(n_2 l_2)
\Lambda_{\kappa}(n l)
\label{A.5}  
\eeq

\noindent
where 

$$
n=\kappa_1 +\kappa_2 +m, \qquad
\Lambda_{\kappa}(n l) \, = \, \frac{ (-)^{\kappa}}{\kappa !}
\pmatrix{ n+l+1/2 \cr n-k \cr}
$$ 

\noindent
(see also Ref. \cite{KVCF}) and

$$
\phi = \left\{ \begin{array}{ l@{\quad \quad} l}
0,  & \kappa - m -n_2 \le 0 \\ 
\kappa - m -n_2 , &  \kappa - m -n_2 > 0 \\ \end{array} \right. , \qquad
\sigma = \left\{ \begin{array}{ l@{\quad \quad} l}
0,  & \kappa - m -\kappa_1 \le 0 \\ 
\kappa - m -\kappa_1 , &  \kappa - m -\kappa_1 > 0 \\ \end{array} 
\right. 
$$

\noindent
For the coefficients $\varepsilon_{\kappa}$ 
used in the present work see Ref. \cite{KVcdm}. 


\newpage
\centerline{\large \bf Figure Captions}

\bigskip
\bigskip
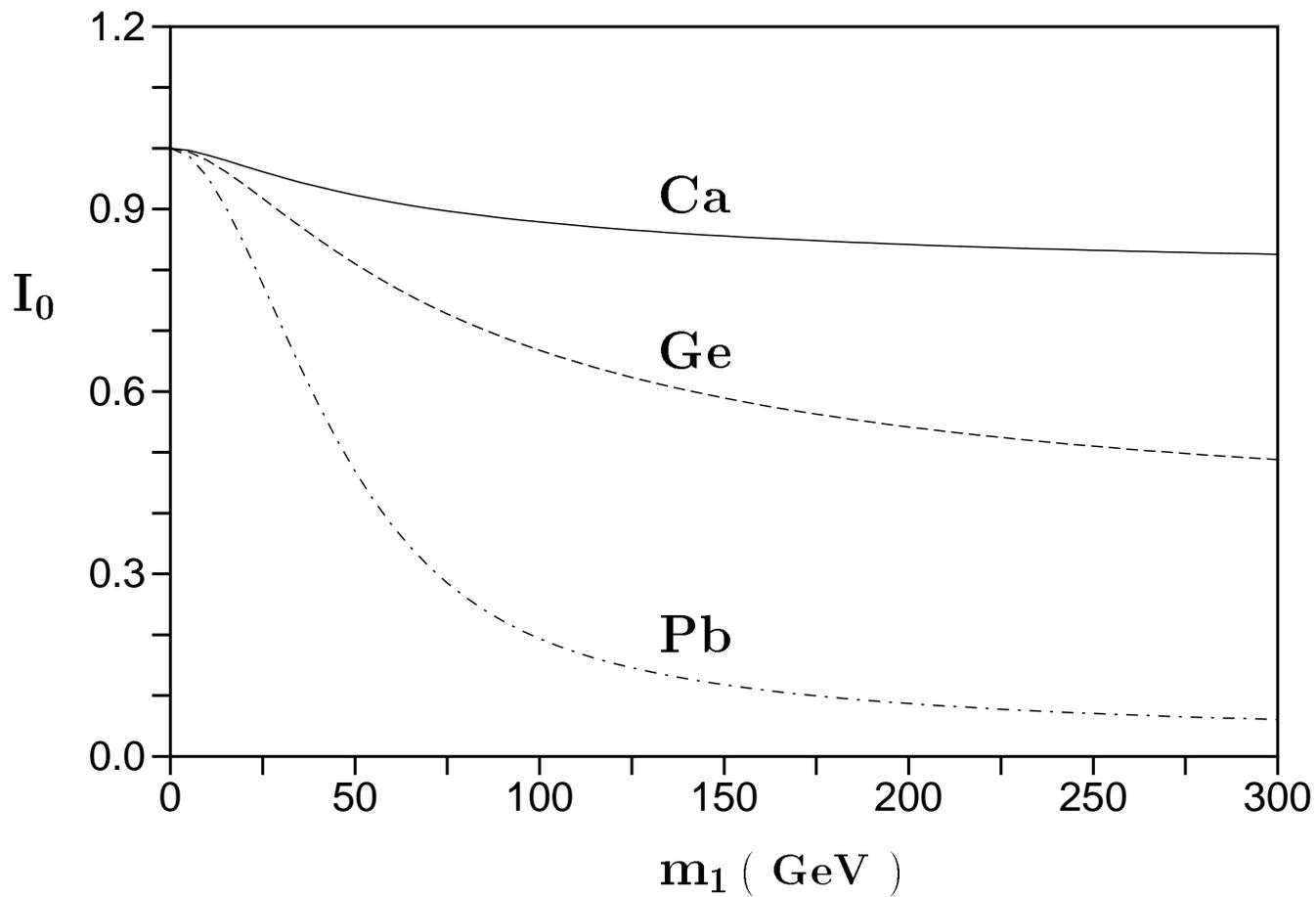
{\bf Fig. 1.}

\bigskip
The integrals $I_0$ (see Eq. (\ref{2.16})), which describe the dominant scalar 
contribution (coherent part) of the total cross section as a function 
of the LSP mass ($m_1$), for three typical nuclei: $Ca$, $Ge$ and $Pb$.
The value $\sqrt{<\beta^2>} =10^{-3}$ was used.

\bigskip
\bigskip
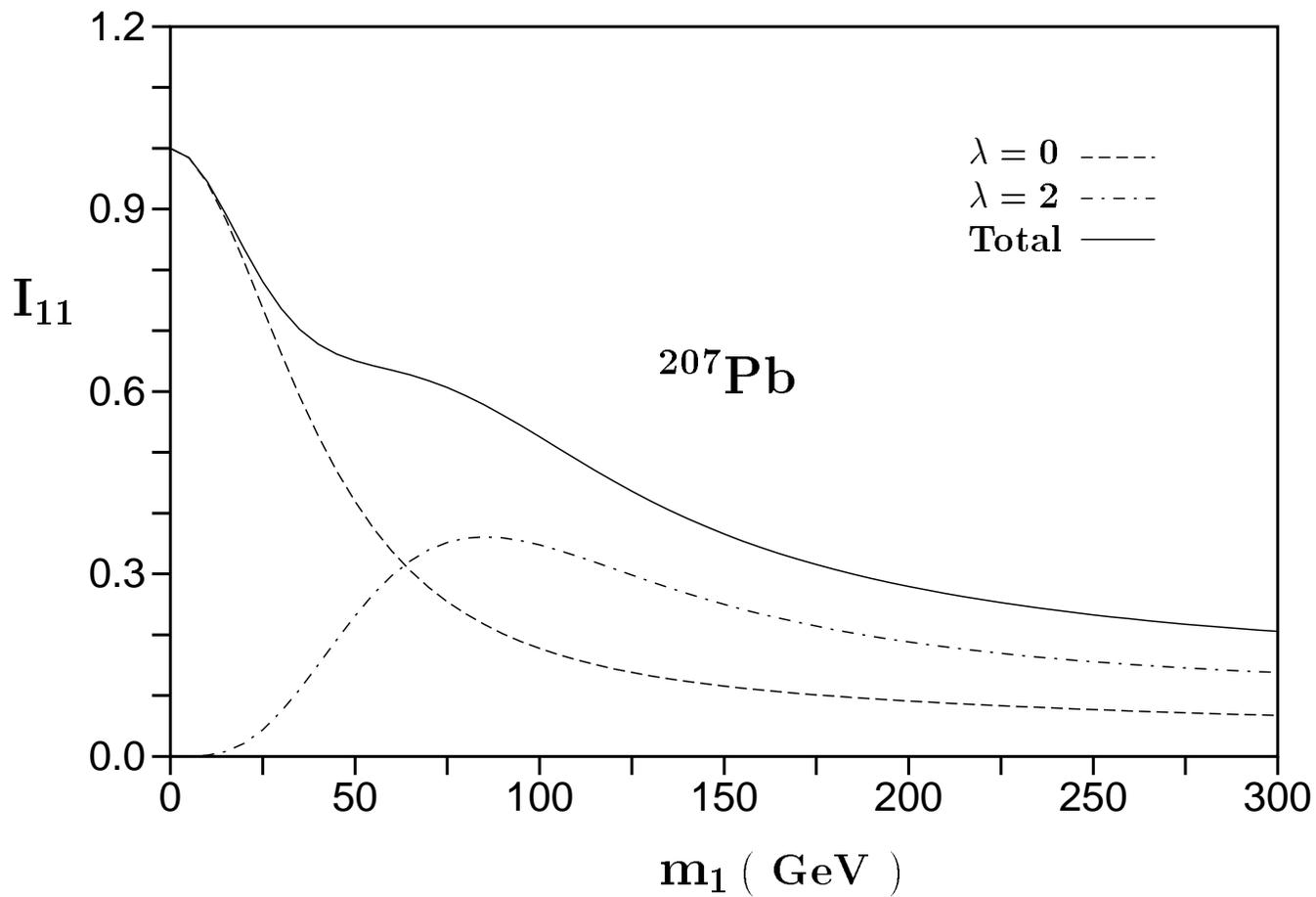
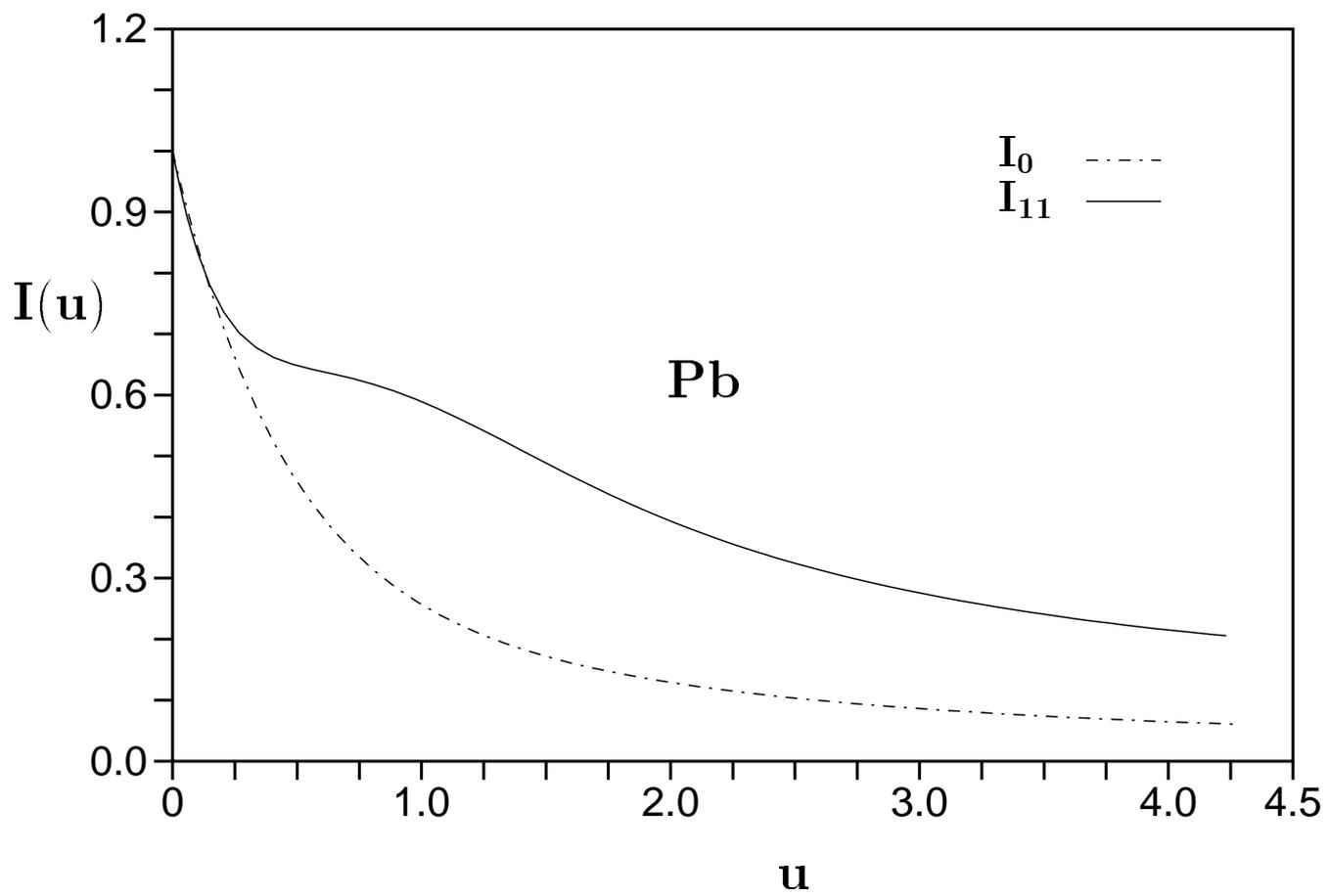
{\bf Fig. 2.}

\bigskip
(a) Plot of the integrals $I_{11}$ as a function of the LSP mass $m_1$.
This integral gives the spin contribution to the LSP-nucleus total 
cross section for $^{207}_{82}Pb$ (for the definition see Eq. (\ref{2.11}))
in the model described in sect. III. 
The integrals $I_{00}$ and $I_{01}$ are similar.
(b) Plot of the integrals $I_{11}(u)$ and $I_0(u)$ for $Pb$, where $u$ is 
given by Eq. (\ref{2.17}). Note that $I_{11}$ is quite a bit less 
retarded compared to $I_{0}$.


\bigskip
\bigskip
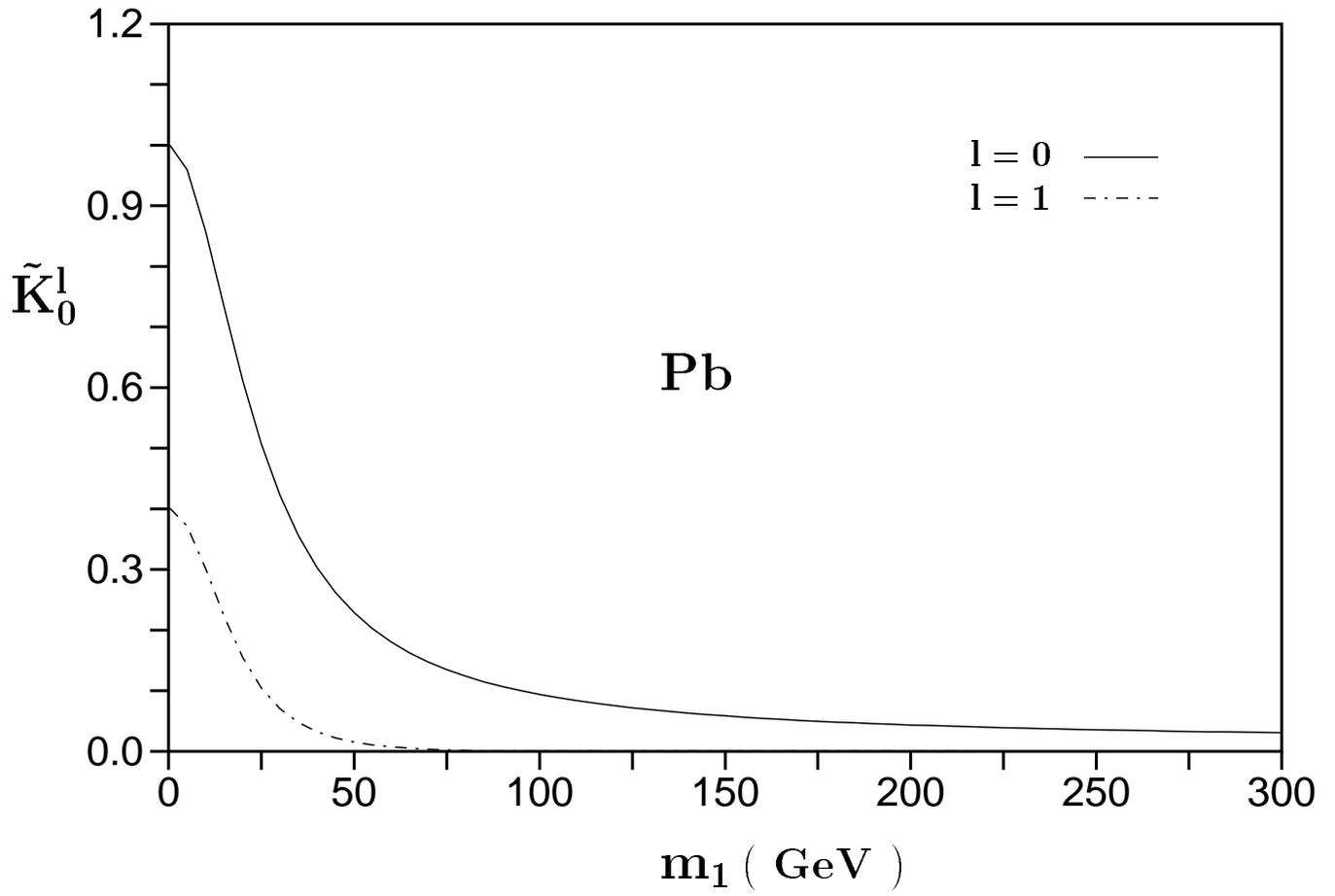
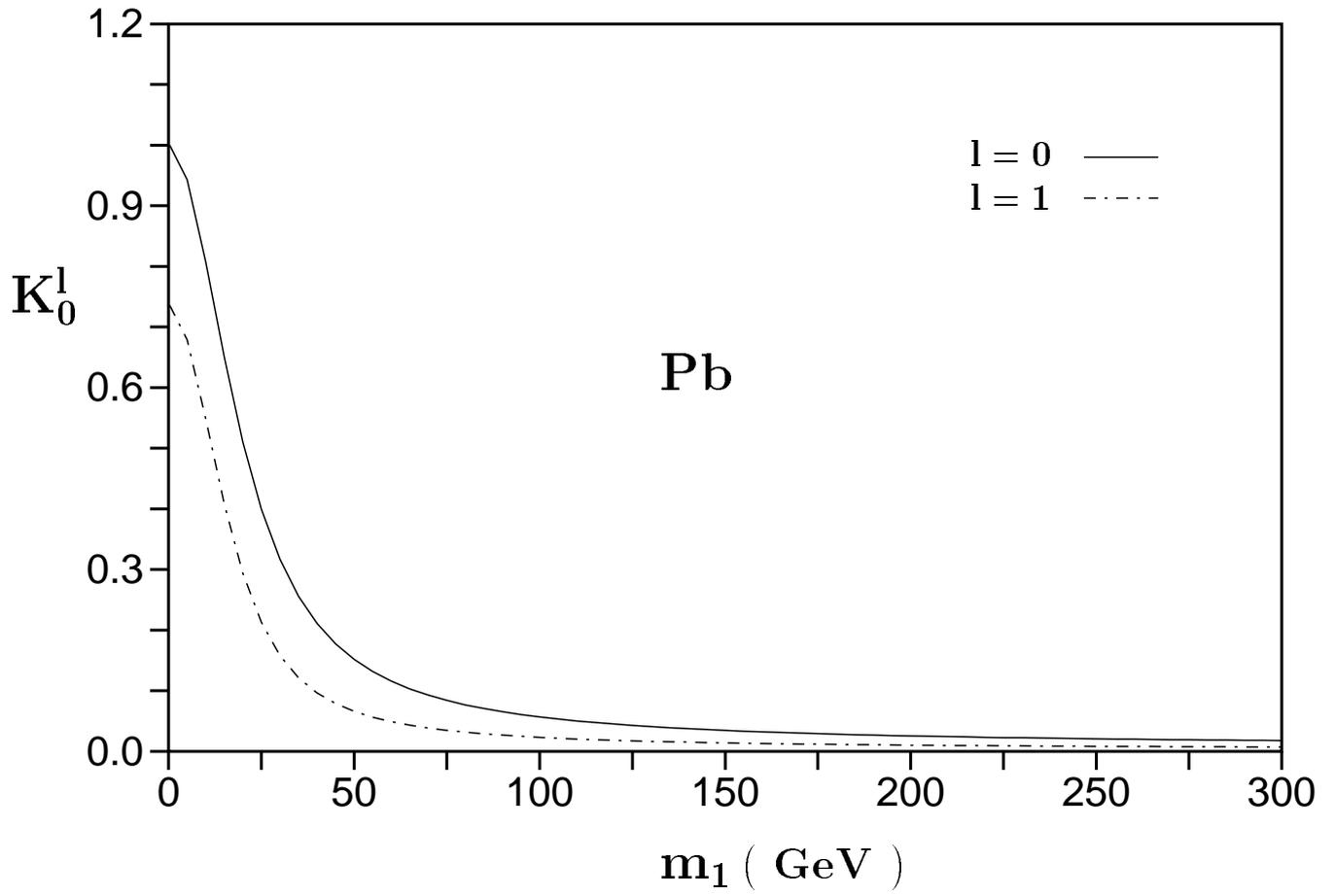
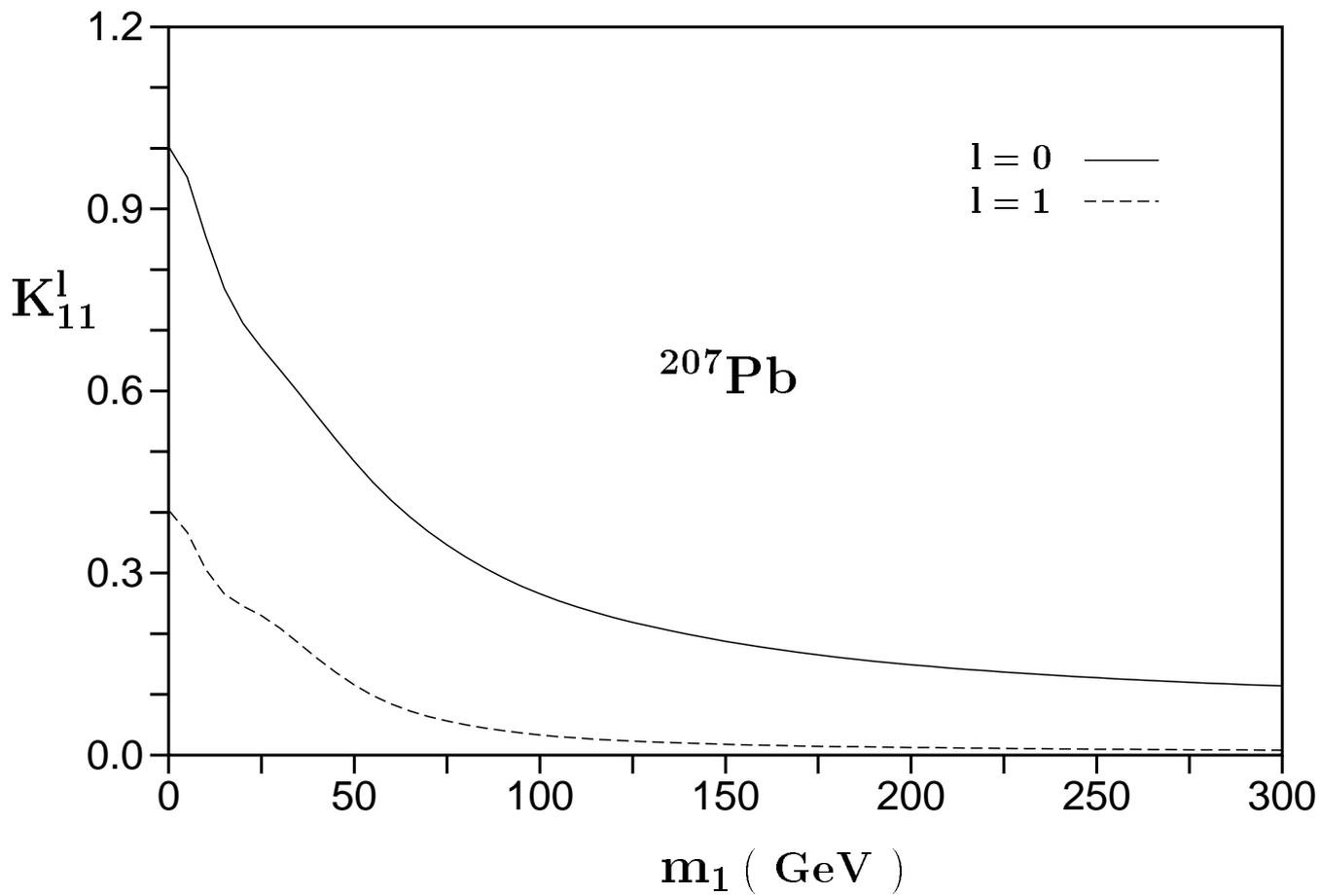
{\bf Fig. 3.}

\bigskip
Contributions of K integrals (for l=0 and l=1) 
defined in Eqs. (\ref{4.30})-(\ref{4.32}) and
entering the event rate due to earth's revolution around the sun:
${\tilde K}^l_0$ in Fig. 3(a), 
$K^l_0$ in Fig. 3(b) and $K^l_{11}$ in Fig. 3(c).
The other integrals $K^l_{00}$ and $K^l_{01}$ are similar to 
$K^l_{11}$. 

\newpage
\begin{table}  
{\bf TABLE I.} The parameters  $\beta f^0_V, f^0_S, f^0_A, f^1_A$ and
$f^1_V/f^0_V, f^1_S/ f^0_S$ for three SUSY solutions (see Ref. \cite{JDV}).
The value of $\beta = 10^{-3}$ was used.  
For the definition of $f^0_S$ and $f^1_S$ (models A, B and C) see Ref. 
\cite{JDV} and for the values of $tan{\beta}$, $m_h$, $m_H$ and $m_A$ employed
see Ref. \cite{Kane}. 

\vspace{0.2cm}

 \begin{tabular}{lccc}
\hline
\hline
 &  &  &  \\
Quantity \hspace{.2cm}& \hspace{.2cm} Solution\hspace{.2cm}$ \#1$ &\hspace{.2cm} Solution
$ \#2$\hspace{.2cm} &\hspace{.2cm} Solution $ \#3$\hspace{.2cm} \\
\hline
 &  &  &  \\
     $\beta f^0_V(Z)$ & $0.475\times10^{-5}$ & $1.916\times10^{-5}$
& $0.966\times10^{-5}$ \\
     $f^1_V(Z)/f^0_V(Z)$ & -1.153 & -1.153 & -1.153\\
     $\beta f^0_V({\tilde q})$ &$1.271\times10^{-5}$&$0.798\times10^{-5}$
 & $1.898\times10^{-5}$ \\
$f^1_V({\tilde q})/f^0_V({\tilde q})$ & 0.222 & 2.727 &0.217 \\
$f^0_s({\tilde q})/\beta f^0_V({\tilde q})(model A)$ &$6.3\times10^{-3}$ & 
$3.6\times10^{-3}$ &$2.4\times10^{-3}$ \\
$f^0_s({\tilde q})/\beta f^0_V({\tilde q})(model B)$ &$0.140$ & 
$3.5\times10^{-2}$ &$5.8\times10^{-2}$ \\
$\beta f^0_V$ &$1.746\times10^{-5}$&$2.617\times10^{-5}$
 & $2.864\times10^{-5}$ \\
$f^1_V/f^0_V$ &-0.153 & -0.113 &-0.251 \\
$LSP$ $mass(GeV)$ &126.0 &27.0 &102.0 \\
$tan{\beta}$ &10.0 &1.5 &5.0 \\
$tan2a$ &0.245 &6.265 &0.528 \\
$m_h,m_H,m_A$ &116,346,345 &110,327,305 &113,326,324 \\
$f^0_S(H)$ (model A) &$1.31\times10^{-5}$&$1.30\times10^{-4}$ &$1.38\times10^{-5}$ \\
$f^1_S/f^0_S$ (model A) &-0.275 &-0.107 &-0.246 \\
$f^0_S(H)$ (model B) &$5.29\times10^{-4}$&$7.84\times10^{-3}$ &$6.28\times10^{-4}$ \\
$f^0_S(H)$ (model C) &$7.57\times10^{-4}$&$7.44\times10^{-3}$ &$7.94\times10^{-4}$ \\
$f^0_A(Z) $ &- &- &- \\
$f^1_A(Z) $ & $1.27\times10^{-2}$&$5.17\times10^{-2}$
 & $2.58\times10^{-2}$ \\
$f^0_A({\tilde q}) (NQM) $& $0.510\times10^{-2}$&$3.55\times10^{-2}$
 & $0.704\times10^{-2}$ \\
$f^0_A({\tilde q}) (EMC) $& $0.612\times10^{-3}$&$0.426\times10^{-2}$
 & $0.844\times10^{-3}$ \\
$f^1_A({\tilde q}) $& $0.277\times10^{-2}$&$0.144\times10^{-2}$
 & $0.423\times10^{-2}$ \\
$f^0_A (NQM) $& $0.510\times10^{-2}$&$3.55\times10^{-2}$
 & $0.704\times10^{-2}$ \\
$f^0_A (EMC) $& $0.612\times10^{-3}$&$0.426\times10^{-2}$
 & $0.844\times10^{-3}$ \\
$f^1_A $& $1.55\times10^{-2}$&$5.31\times10^{-2}$
 & $3.00\times10^{-2}$ \\ 
\hline
\hline
\end{tabular}
\end{table}

\vspace{.6cm}

\begin{table}  
{\bf TABLE II.} The quantity $q_0$ (forward momentum transfer) in units of
$fm^{-1}$ for three values of $m_1$ and three typical nuclei.
In determining $q_0$ the value $\sqrt{<\beta^2>} =10^{-3}$ was employed
(see Eq. (\ref{2.8})).

\vspace{0.2cm}

\begin{tabular}{cccc}
\hline
\hline
  &  &  &  \\
& \multicolumn{3}{c}{ $q_0$ ($fm^{-1}$)}  \\
\hline
  &  &  &  \\
Nucleus & $m_1=30.0 \, GeV$ & $m_1=100.0 \, GeV$ & $m_1=150.0 \, GeV$ \\
\hline
  &  &  &  \\
  $Ca$  &  0.174   &  0.290 &  0.321 \\
  $Ge$  &  0.215   &  0.425 &  0.494 \\
  $Pb$  &  0.267   &  0.685 &  0.885 \\ 
\hline
\hline
\end{tabular}
\end{table}

\newpage
\begin{table}  
{\bf TABLE III.} The quantity $\Big<\frac{dN}{dt}\Big>_0$ in $y^{-1}Kg^{-1}$ 
and the
parameter h (oscillation due to the earth's motion around the sun) for 
the coherent vector and scalar contributions. For the definition of A, B, C,
see text. NFF and WNFF stand for Nuclear Form Factor and Without Nuclear 
Form Factor, respectively.

\vskip0.2cm

\begin{tabular}{rlcclrrc}
\hline
\hline
& & & & & & &   \\
 & & \multicolumn{2}{c}{Vector \hspace{.2cm} Contribution}  &
     \multicolumn{4}{c}{Scalar \hspace{.2cm} Contribution}  \\
\hline 
& & & & & & &   \\
& & $\Big<\frac{dN}{dt}\Big>_0 (y^{-1} Kg^{-1})$ &$ h $ &
     \multicolumn{3}{c}{$\Big<\frac{dN}{dt}\Big>_0 (y^{-1} Kg^{-1})$}
     & $h$ \\ 
\hline
& & & & & & &   \\
& \multicolumn{1}{c}{Solution} &  &  &
    \multicolumn{1}{c}{ Model \hspace{.2cm} A }& 
    \multicolumn{1}{c}{ Model \hspace{.2cm} B }& 
    \multicolumn{1}{c}{ Model \hspace{.2cm} C }&  \\ 
\hline
& & & & & & &   \\
   &  $\#1 (NFF) $  & $0.264\times 10^{-3}$ & 0.029 & $ 0.151\times 10^{-3}$ & 
   0.220 & 0.450 & -0.002  \\
   &  $\#1 (WNFF) $  &$0.378\times 10^{-2}$ & 0.102 & $ 0.212\times 10^{-2}$ & 
   3.087 & 6.322 & 0.054  \\
   &  $\#2 (NFF) $  & $0.162\times 10^{-3}$ & 0.039 & $0.410\times 10^{-1}$ &
   142.860 & 128.660 & 0.026  \\
Pb &  $\#2 (WNFF) $  & $0.332\times 10^{-3}$ & 0.102 & $0.872\times 10^{-1}$ & 
   303.390 & 273.220 & 0.054  \\
   &  $\#3 (NFF) $  & $0.895\times 10^{-3}$ & 0.038 &  $0.200\times 10^{-3}$ & 
   0.377 & 0.602 & -0.001  \\
   &  $\#3 (WNFF) $  & $0.970\times 10^{-2}$ & 0.102 &  $0.220\times 10^{-2}$ &
   4.114 & 6.576 & 0.054  \\
\hline
& & & & & & &   \\
   &  $\#1 (NFF) $  & $0.151\times 10^{-3}$ & 0.043 &  $0.779\times 10^{-4}$ &
   0.120 & 0.245 & 0.017  \\
   &  $\#1 (WNFF) $  & $0.518\times 10^{-3}$ & 0.102 &  $0.230\times 10^{-3}$ 
   & 0.353 & 0.723 & 0.054  \\
Ge &  $\#2 (NFF) $  & $0.527\times 10^{-4}$ & 0.057 &  $0.146\times 10^{-1}$ &
  51.724  & 46.580 & 0.041  \\
   &  $\#2 (WNFF) $  & $0.708\times 10^{-4}$ & 0.102 &  $0.196\times 10^{-1}$ &
   69.506 & 62.595 & 0.054  \\
   &  $\#3 (NFF) $  & $0.481\times 10^{-3}$ & 0.045 &  $0.101\times 10^{-3}$ &
   0.198 & 0.316 & 0.020  \\
   &  $\#3 (WNFF) $  & $0.144\times 10^{-2}$ & 0.102 &  $0.267\times 10^{-3}$ &
   0.525 & 0.839 & 0.054  \\
\hline
& & & & & & &   \\
   &  $\#1 (NFF) $  & $0.790\times 10^{-4}$ & 0.053 &  $0.340\times 10^{-4}$ &
   0.055 & 0.114 & 0.037  \\
   &  $\#1 (WNFF) $  & $0.135\times 10^{-3}$ & 0.102 &  $0.520\times 10^{-4}$ 
   & 0.085 & 0.174 & 0.054  \\
Ca &  $\#2 (NFF) $  & $ 0.264\times 10^{-3}$ & 0.060 &  $0.612\times 10^{-2}$ &
   22.271 & 20.056 & 0.048  \\
   &  $\#2 (WNFF) $  & $0.307\times 10^{-3}$ & 0.102 &  $0.704\times 10^{-2}$ &
   25.601 & 23.055 & 0.054  \\
   &  $\#3 (NFF) $  & $0.241\times 10^{-3}$ & 0.053 &  $0.435\times 10^{-4}$ &
   0.090 & 0.144 & 0.038  \\
   &  $\#3 (WNFF) $  & $0.388\times 10^{-3}$ & 0.102 &  $0.642\times 10^{-4}$ &
   0.133 & 0.212 & 0.054  \\
\hline
\hline
\end{tabular}
\end{table}

\vspace{0.6cm}

\begin{table}  
{\bf TABLE IV.} The spin contribution in the LSP-nucleus scattering in 
$^{207}Pb$ for two cases: EMC data and NQM Model (see sect. II). For 
other notations see caption of Table III.

\vskip0.2cm

\begin{tabular}{llllc}
\hline
\hline
& & & &   \\
 & \multicolumn{2}{c}{\hspace{1.2cm}EMC \hspace{.2cm} DATA} \hspace{.8cm} &
   \multicolumn{2}{c}{\hspace{1.2cm}NQM \hspace{.2cm} MODEL}\hspace{.8cm} \\ 
\hline
& & & &   \\
Solution & \hspace{.2cm}$\Big<\frac{dN}{dt}\Big>_0 $ $(y^{-1} Kg^{-1})$
    \hspace{.2cm} & $ h $ & 
	 $\hspace{.2cm} \Big<\frac{dN}{dt}\Big>_0 $ $ (y^{-1} Kg^{-1})$ & $ h $ \\ 
\hline
& & & &   \\
$\#1 (NFF) $  &$0.285\times 10^{-2}$& 0.014 &$0.137\times 10^{-2}$& 0.015  \\
$\#1 (WNFF)$  &$0.130\times 10^{-1}$& 0.054 &$0.551\times 10^{-2}$& 0.054  \\
$\#2 (NFF) $  & 0.041               & 0.046 &$0.384\times 10^{-2}$& 0.056  \\
$\#2 (WNFF)$  & 0.062               & 0.054 &$0.405\times 10^{-2}$& 0.054  \\
$\#3 (NFF) $  & 0.012               & 0.016 &$0.764\times 10^{-2}$& 0.017  \\
$\#3 (WNFF)$  & 0.047               & 0.054 &$0.269\times 10^{-1}$& 0.054  \\
\hline
\hline
\end{tabular}
\end{table}

\end{document}